\documentclass[floatfix,altaffilletter,showpacs,twocolumn]{revtex4}
\usepackage{graphicx,float,amsmath}

\begin{document}


\title{Effect of premelting on conductivity of DNA-lipid films}

\author{A.Yu.~Kasumov} 
\email[To whom correspondence should be addressed. E-mail: ]{kasumov@lps.u-psud.fr}
\affiliation{Laboratoire de Physique des Solides, Associ\'e au CNRS, B\^atiment 510, 
Universit\'e Paris-Sud, 91405 Orsay, France}
\affiliation{Institute of Microelectronics Technology and High Purity Materials, RAS, 
Chernogolovka 142432 Moscow Region, Russia}
\affiliation{RIKEN, Hirosawa 2-1, Wako, Saitama, 351-0198 Japan}
\author{S.~Nakamae}
\altaffiliation[Present address: ]{Service de Physique de l'Etat Condens\'e, CEA-Saclay, 
91191 Gif-sur-Yvette, France}
\affiliation{Laboratoire de Physique du Solide (UPR 5 CNRS) ESPCI, 10 rue Vauquelin 75231 Paris, France}
\affiliation{Laboratoire de Mat\'eriaux et Ph\'enom\`ene Quantiques (UMR 7162) Universit\'e Denis-Diderot 
(Paris 7) Paris, France}
\author{M.~Cazayous}
\affiliation{Laboratoire de Physique du Solide (UPR 5 CNRS) ESPCI, 10 rue Vauquelin 75231 Paris, France}
\affiliation{Laboratoire de Mat\'eriaux et Ph\'enom\`ene Quantiques (UMR 7162) Universit\'e Denis-Diderot 
(Paris 7) Paris, France},
\author{T. Kawasaki}
\affiliation{Department of Biomolecular Engineering, Tokyo Institute of Technology, 4259 Nagatsuda, 
Midori-ku, Yokohama 226-8501, Japan}
\author{Y. Okahata}
\affiliation{Department of Biomolecular Engineering, Tokyo Institute of Technology, 4259 Nagatsuda, 
Midori-ku, Yokohama 226-8501, Japan}

\begin{abstract}
We have measured temperature dependent (between 20 and 80 $^\circ$C) electrical conductivity and 
molecular structure (Raman spectroscopy) of DNA-lipid cast film. 
Our findings show that the conductivity is strongly influenced by premelting effects 
in the molecular structure starting near physiological temperatures ($\sim$40 $^\circ$C), 
prior to the global DNA denaturation.
\end{abstract}

\maketitle

Most measurements reported in the last decade on the DNA conductivity are 
conducted at room temperatures and below \cite{porath}. 
If DNA is to become exploitable in micro-electronics applications, 
however, its performance must be reliable at temperatures 
slightly {\it above} the room temperature due to the inevitable heating of electronic 
components. It is well known that DNA molecules, both natural and synthetic, undergo a
denaturation process at $T_{dn}$ = 70 $\sim$ 80 $^\circ$C. 
Above this temperature, the double-stranded molecular conformation is destroyed, and consequently, 
the electrical conductivity is lost \cite{okahata1, Iqbal}.
According to numerous theoretical models \cite{hwa, Peyrard} even at physiological 
temperatures ($\sim$40 $^\circ$C), DNA experiences structural 
perturbations leading to local denaturations and/or ``bubble"-type defect formations. 
The existence of ``bubbles'' \cite{Strick, Schallhorn} as well as the temperature induced 
local perturbations at $T< T_{dn}$, termed ``premelting'' 
\cite{Movileanu, Movileanu2, Mukerji, Carrier, Tomlinson} has been confirmed experimentally, 
and the biological aspects of these local denaturations were discussed in a number of 
studies (see for example, \cite{Strick}). 
Local deformations should cause breaking of a 
long-range order in the DNA structure ({\it i.e.}, interruption of the parallel 
base-pair (bp) stackings) similar to an order breaking in solid bodies due to the 
dislocation introduction.  
But their influence on conductivity has not been properly addressed until now. 
In this letter, we present the temperature dependent conductivity and structural 
evolution monitored through Raman spectroscopy measured on the DNA-lipid cast film 
between physiological and denaturation temperatures. 
These DNA-lipid films were previously studied by Okahata {\it et al.} \cite{okahata1}, 
where disappearance of conductivity above the denaturation temperature was reported. 
We observed a substantial reduction in the DNA conductivity due to premelting effects 
starting at temperature as low as 40 $^\circ$C, lending support to the theoretical inference 
on the importance of the long range parallel $bp$ stacking in DNA for the electrical conduction.

\begin{figure}[!htp] 
	\begin{center} 
	\includegraphics[width=8cm]{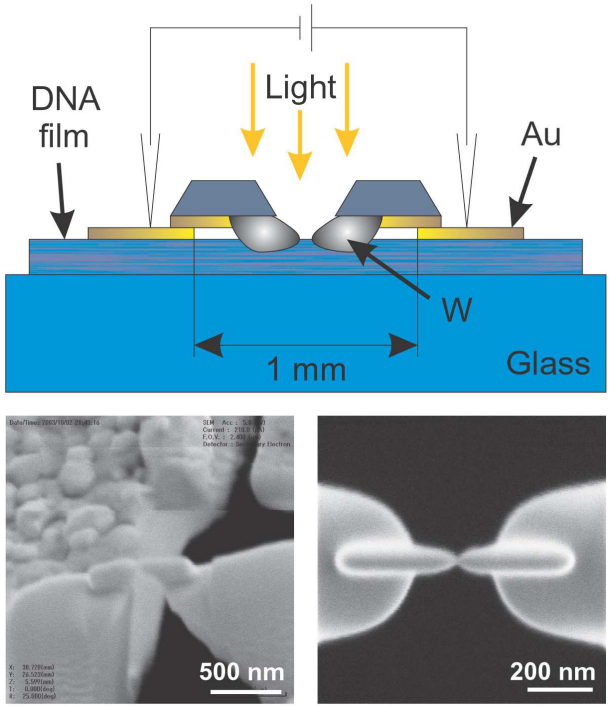}
	\end{center}
\caption{Conductivity measurement set-up. One micron high tungsten nanoelectrodes with a less than 5 nm 
separation were prepared by decomposing Tungsten hexacarbonyl by focused ion beam 
\cite{method} on a silicon chip. The silicon chip was then pressed into the DNA-lipid film surface 
on which gold contacts had been pre-deposited. Below are
SEM (left,side view) and FIB (right,top view) images of the tungsten nanoelectrodes.}
\label{fig.1}
\end{figure}

Self-standing DNA-lipid cast films with thickness of about 60 microns were prepared according 
to the method described elsewhere \cite{okahata2}. 
Once the self-standing film is mechanically 
stretched, DNA molecules (natrual DNA, 2000 $bp$'s each) are aligned along film's long axis 
with an average inter-molecular distance of 41 \AA. 
In previous investigations, Okahata {\it et al.} have verified the anisotropic 
conductivity through these films and concluded that the electrons are traveling 
through the molecules and not via the lipid matrix (conductivity of these molecules was not 
suppreseed by strong interaction with a solid surface \cite{r14}).  
For our conductivity measurements, a section of a film of about 5x0.5 cm$^2$ was placed on a glass plate.  
The detailed description of measurement apparatus is given in Figure~\ref{fig.1}. 
The measurements were performed in a Hewlett-Packard 
measurement systems in a dark box at temperatures of 30-150 $^\circ$C and 
with or without illumination. 
The leak current through the contacts was less than 1 nA in the measurement 
range of $\pm$ 3 V and for temperatures between 25 and 80 $^\circ$C, 
irrespective of illumination. 
Indentation created by the electrodes in soft insulators, 
such as resist or teflon, leads only to a reduction of the leak current 
due to an increase in the distance between electrodes. 
Similar increase is expected in the DNA film, while the total current ($I$) at 3 V reaches 1 $\mu$A 
and larger. Once heating the film to 150 $^\circ$C, $I$ again decreases down to $\sim$ 1 nA.

\begin{figure}[!htp] 
	\begin{center} 
	\includegraphics[width=8cm]{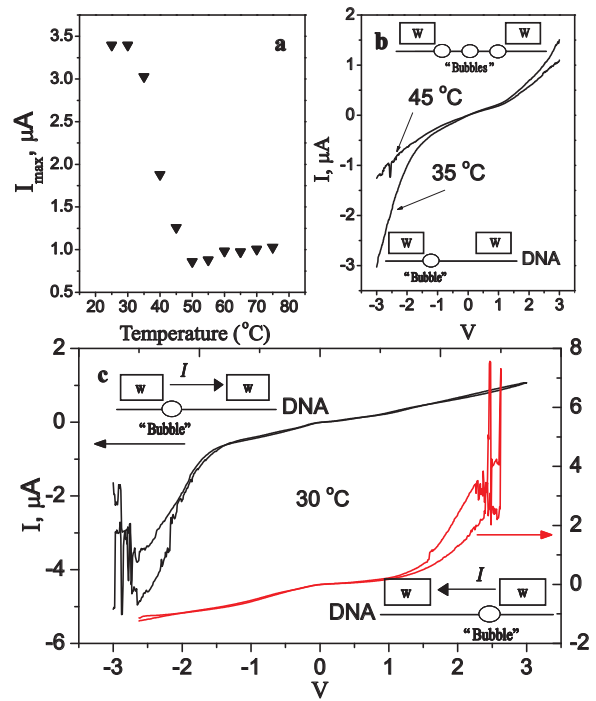}
	\end{center}
\caption{Temperature dependent conductivity measurements under luminosity. 
a) Temperature dependence of $I_{max}$, 
b) $I-V$ characteristics (at 30 $^\circ$C) of the DNA film at 35 and 45 $^\circ$C. The insets show the
increasing number of bubbles with a temperature, 
c) Irreversible $I-V$ characteristics of the DNA film which had been subjected to a high bias voltage 
at 80 $^\circ$C with opposing polarities.  The insets show the assumed schematics of
the bubble type defect movements inside the DNA molecules at 80 $^\circ$C.}
\label{fig.2}
\end{figure}

With illumination of 3 mW/cm$^2$ (Halogen photo optic lamp, unfiltered and unpolarized), 
the overall conductivity and the maximal current ($I_{max}=I(-3V)$) 
through the DNA-film start to diminish at about 40 $^\circ$C (Figure~\ref{fig.2}a). 
The largest change is observed between 35 and 45 $^\circ$C (Figure~\ref{fig.2}b) where $I_{max}$ is reduced by 
more than 70 \%. The stability of the temperature was $\pm$ 0.5 $^\circ$C during I-V characteristics 
measurements. When measurements are conducted in dark, the conductivity is 5 times smaller at
35 $^\circ$C and the temperature dependent reduction of $I_{max}$ is limited to 20 \% (not shown).
These temperature induced changes in the conductivity are entirely reversible upon thermal
cycling provided that the DNA-film is not subjected to a high voltage bias ($\pm$ 3V) 
at the highest temperature (80 $^\circ$C). 
The most probable reason for the conductivity reduction is the creation of
local defects (premelting), as indicated by Raman spectrum evolution (described below),
analogous to that of local dislocations in a solid body during high temperature annealing.
In the case of a solid body, local dislocations can physically migrate at elevated temperatures
and stop close to grain or phase boundaries \cite{mobile}. 
Therefore local defects in DNA can be expected to also advance close to the tungsten nanoelectrodes,
inducing irreversible changes in the molecule-metal contact characteristics \cite{electric}. 
We indeed observe such irreversibility in conductivity when the sample was maintained at 80 $^\circ$C for 
10-15 minutes with the maximum voltage (3V) of a chosen polarity. 
After cooling the sample back down to 30 $^\circ$C, 
I(V) instability is observed on the positive branch (Figure~\ref{fig.2}c). 
When the sample was re-heated to 80 $^\circ$C with $V$ of the opposite sign, 
the instability appears on the negative branch of I(V) curve after cooling to 
30$^\circ$C. We observed such behavior up to 4 cycles of measurements.

In order to associate the conductivity changes with DNA structural modifications, 
we have tracked the structural evolution of DNA molecules by Raman spectroscopy. 
The 514.5 nm excitation line of an Ar$^+$-Kr$^+$ laser was focused on the samples through a $\times$50 
magnification objective lense with a radiation power at source of 10 mW. 
The scattered light was analysed using a Jobin-Yvon triple grating spectrometer in the confocal configuration. 
The effective spectral resolution was less than 1 cm$^{-1}$. 
Raman spectra were taken in the 10-80$^\circ$C temperature range on several films from the same batch as the 
conductivity measurements. The spectra of DNA-lipid complex taken at room temperature and at 80
$^\circ$C as well as that of natural B-DNA (also taken in our lab) are compared in Figure~\ref{fig.2}.
Temperature dependent Raman sepctra of the DNA-lipid complex films showed certain notable differences 
from those observed in DNA molecules in aquaous solution. 
First, a cooperative melting of double-stranded DNA was not observed even at
80$^\circ$C, whereas the pre-melting effects in 10-65$^\circ$C range were clearly present. 
We also remarked that nearly all vibrational modes exhibit a reversible temperature dependency 
during premelting after the heating and the subsequent cooling of the films (not shown). 
Structural rigidity of molecules imposed by the lipid intercalation may explain these differences.
It should also be noted that the hypochromic effect \cite{Tomlinson} was not observed with exception of 
1680cm$^{-1}$ marker (see discussion below) due to the progressive change in the background luminescence 
of the film. Therefore, we have used the heating induced shifts in Raman peaks to track the
premelting effects in DNA molecules. These peaks are sensitive to 
i) backbone and deoxynucleoside conformations, ii) interbase hydrogen bonding and iii) base stacking effect. 

\begin{figure}[!htp] 
	\begin{center} 
	\includegraphics[width=8cm]{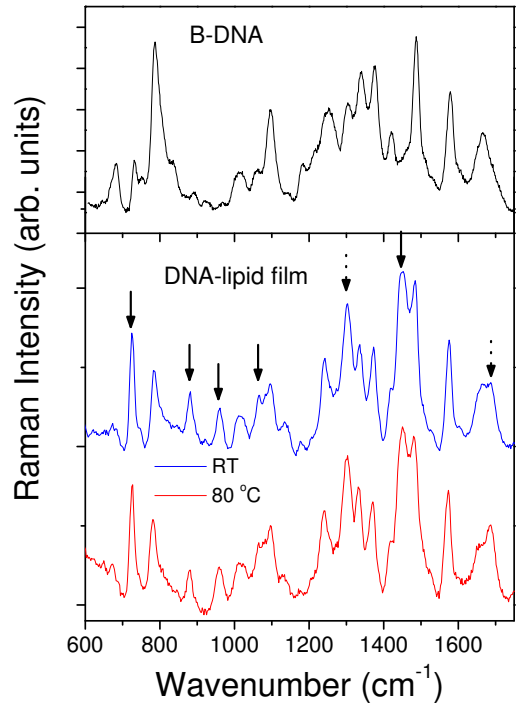}
	\end{center}
\caption{Raman sepctra of natural B-DNA (top) and the DNA-lipid complex at room temerature and at 80$^\circ$C.
The solid arrows indicate the Raman peaks unique to the lipid complex. The dotted arrows indicated the peaks that contain
contributions from both DNA molecules and the lipid complex. For detailed annotation of individual Raman bands, the readers
are kindly asked to refer to literature, for example, \cite{Movileanu},\cite{Deng} and \cite{Duguid}. The Raman bands 
corresponding to the lipid-complex did not show heating induced shifting while certain DNA bands moved to lower wavenumbers.
The DNA PO$_2^-$ symmetric stretching mode remains at 1092cm$^{-1}$ through out the measurements indicating that 
the DNA molecules remains in their B-form without a signigicant change in the relative humidity around the molecues.}
\label{fig.3}
\end{figure}
\begin{figure}[!htp] 
	\begin{center} 
	\includegraphics[width=8cm]{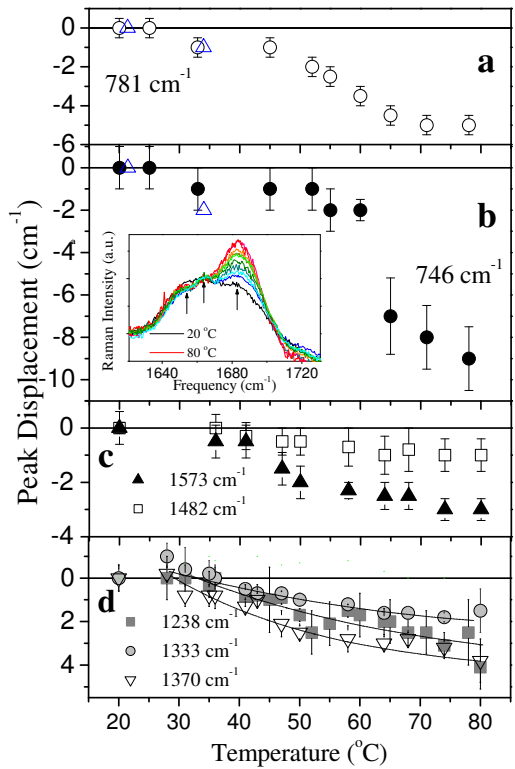}
	\end{center}
\caption{Temperature dependent pre-melting effect in Raman frequencies of {\bf a:} 
The admixture of complex vibration of the B-DNA phosphodiester backbone 
conformation and thymine and cytosine ring modes \cite{Deng}, 
{\bf b:} dT C2'-{\it endo} sugar pucker and {\it anti} glycosyl torsion mode \cite{Movileanu} 
and {\bf c:} dG and dA ring stretching vibrations sensitive to hydrogen bonding \cite{Deng, Duguid} 
as well as dT \cite{Mukerji}.
The bands shift toward lower frequencies starting near 35 $^\circ$C and the movement attenuates 
for 65 $^\circ$C and higher for phosphodiester geometry and deoxyribose hydrogen bond vibrations.
The onset of pre-melting effect at 746 cm$^{-1}$ occurs near 50 $^\circ$C. {\bf d:} Interbase hydrogen
bonding. The bands returns to their original position 
once the film is cooled back down to room temperature (see blue triangles in panels {\bf a} and 
{\bf b}). {\bf inset b:} The intensity change near 1660-1680 cm$^{-1}$ are observable from 35 $^\circ$C and higher 
with no sign of attenuation. 
The spectra are normalized to the intensity values at 1664 cm$^{-1}$ in order to show
the relative intensity change among three bands (indicated by 3 arrows). 
The continued enhancement of 1680 cm$^{-1}$ from room temperature to 80 $^\circ$C 
is clearly observed. The bands at 1664 and 1650 cm$^{-1}$ are indistinguishable at 20 $^\circ$C 
(thick line) separate themselves into two distinct peaks at 1664 and 1648 cm$^{-1}$ at 
80 $^\circ$C (thin line).}
\label{fig.4}
\end{figure}


i) The broad peak at 780 cm$^{-1}$ contains contributions from dC, dT and the 5'C-O-P-O-C3' backbone stretching. 
The peak frequency shifts down from 781 to 776 cm$^{-1}$) starting around 45$^\circ$C until near 70$^\circ$C 
(see Fig.~\ref{fig.4}a). The peak located at 840 cm$^{-1}$ originates from the 5'C-O-P-O-C3' phosphodiester backbone 
movements and can be used as a quantitative measure of the ordered phosphodiester. 
This peak decreases in intensity and disappears into the background. The 746 cm$^{-1}$ peak is the vibrational 
marker of C2'-endo/anti conformation of dT. This peak broadens and shifts strongly to lower frequency indicating 
the extended distribution of conformations at higher temperature (Fig.~\ref{fig.4}b).  
The onset of this movement is 55$^\circ$C and continues to shift toward lower frequency up to 80$^\circ$C. 

ii) The bands at 1482 and 1573 cm$^{-1}$ correspond to ring stretching vibrations of purine imidazole ring 
and are sensitive to hydrogen bonding. The both peaks shift to lower frequencies by 1 and 3 cm$^{-1}$, 
respectively, between 35 and 40$^\circ$C and stabilize for temperature above 65$^\circ$C (Fig.~\ref{fig.4}c). 
These shifts are the signature of the thermo-instability of base pairing. Moreover, the 3 cm$^{-1}$ 
shift down of the Raman peaks between 1200-1400 cm-1 is associated with elimination of hydrogen bonding 
between bases (Fig.~\ref{fig.4}d). 

iii) The intensity evolution of three peaks located at 1658, 1668 and 1682 cm$^{-1}$ (inset of Fig.~\ref{fig.4}b) 
are normally attributed to the base stacking effects related to carbonyl stetching vibrations 
coupled to ring stretching vibration (mostly dT) \cite{Duguid}. The Raman peak at 1682cm$^{-1}$ of the DNA-lipid film is; 
however, considerably more intense than in a typical B-DNA (see Fig~\ref{fig.3}), 
suggesting that contribution from the lipid complex cannot be ignored. 
Hence we cannot conclude the hypochormicity observed here to the effect of DNA pre-melting without further investigation. 

As described above, the premelting effects observed in DNA Raman isgnatures coincide with the
temperature dependence of electrical conductivity measured in these films.  
Local destruction of double-stranded DNA conformation caused by
premelting effects is most simply described as a 
``bubble'' creation.  
The formation of such bubbles, and more importantly, its movement
within the molecule is just like a dislocation loop moving 
in a metal microwire subjected to a high current density \cite{microwire}. The insets in fig.~\ref{fig.2}c will
help understand the observed changes in electrical conductivity (Fig.~\ref{fig.2}). 
Probably, in the presence of defects, the potential barrier in the vicinity of the 
molecule-metal contact decreases. 
This barrier reduction is a known effect in metal-semiconductor microcontacts 
near a dislocation \cite{semicon} and electromigration, the physical displacement of 
defects under an applied electrical current, is also well-known \cite{electromigration}.  
In electromigration, the direction of dislocation movement can be switched by changing the 
current direction. 
To estimate the pressure on a ``bubble'' by an electron wind, the
current density in a DNA molecule must be known. 
Emerging consensus states that DNA molecules 
longer than 10 nm combined with bad electric contacts become insulating \cite{contacts}. 
Therefore, only about 10 molecules at the film surface should be electrically active 
in our measurements. 
Taking 2 nm as the DNA diameter, the corresponding current density would be as high as 10$^7$ A/cm$^2$, 
comparable to a current density required to drag a dislocation in a metal microwire \cite{microwire}. 
One can estimate the applied force on a bubble, $F_b$, using the equation
for a dislocation in a metal \cite{density}: $F_b = jm^{*}V_fS_b/e$ ,  where $j$ is the 
current density, $m^*$ and $V_f$ are the effective mass and Fermi velocity of 
the electrons injected in DNAs from the tungsten electrods \cite{23}, $S_b$ is 
the scattering cross section of the electrons by a bubble, and $e$ 
is the electron charge. The applied force is about 1~pN (with $m^* \sim$  
10$^{-30}$ kg and $V_f \sim$ 10$^6$ m/s in W \cite{24}, $S_b \sim$ 1nm$^2$ for a small bubble) 
which is enough to deform a DNA molecule \cite{Strick}.
The number of bubbles increases with a temperature \cite{Peyrard}, and it decreases the 
conductivity of DNAs (Fig.~\ref{fig.2}.b) and makes I-V characteristics more 
symmetrical (probably due to symmetrical molecule-metal contacts, see 
insets in Fig.~\ref{fig.2}b).

In summary, our measurements demonstrate that the electrical conduction in DNA 
can be compromised under a moderate heating above room temperature
due to local disruptions in the long-range B-DNA structure.  
Furthermore, displacement of defects along molecules could 
explain why the structural transformation, as probed by Raman spectroscopy, is a reversible
process while the electrical conductivity is not.  

We thank D.Klinov, H.Bouchiat, S.Gu\'eron, A.Braslau and K.Tsukagoshi for useful discussions and
acknowledge the financial support by the Russian Foundation for Basic Research and ANR Quantadn.

\end{document}